\documentclass[fleqn,10pt]{wlscirep}
\usepackage[utf8]{inputenc}
\usepackage[T1]{fontenc}
\usepackage{graphicx}
\usepackage{placeins}
\usepackage{amsmath}
\usepackage{xcolor}

\title{Strategies for COVID-19 vaccination  under a shortage scenario: a geo-stochastic modelling approach.}

\author[a,*]{N. L. Barreiro}
 \affil[a]{ Instituto de Investigaciones Científicas y Técnicas para la Defensa (CITEDEF), 1603, Buenos Aires, Argentina.
}
 
\author[b]{C. I. Ventura}
\affil[b]{ (CONICET) Centro  Atómico Bariloche-CNEA, and Universidad Nacional de Río Negro, 8400-Bariloche, Argentina}

\author[c]{T. Govezensky}
\affil[c]{ Instituto de Investigaciones Biomédicas, Universidad Nacional Autónoma de México, México 04510, México.
 }%

\author[d,e,f]{M. Núñez}
\affil[d]{Consejo Nacional de Investigaciones Científicas y Técnicas (CONICET), Buenos Aires, Argentina}
\affil[e]{Departamento Materiales Nucleares, Centro Atómico Bariloche,
Comisi\'on Nacional de Energía Atómica (CNEA), Bariloche, Argentina}
\affil[f]{ INIBIOMA ,Universidad Nac. del Comahue,Bariloche, Argentina
}

\author[g,a]{P. G. Bolcatto}
\affil[g]{ Instituto de Matemática Aplicada del Litoral (IMAL, CONICET/UNL), FHUC. Santa Fe,3000, Argentina}

\author[h]{R. A. Barrio}
\affil[h]{ Instituto de Física. Apartado Postal 20-365, Universidad Nacional Autónoma de México, México 04510, México.}

\affil[*]{nadus.barreiro@gmail.com}

\begin{abstract}
In a world being hit by waves of COVID-19, vaccination is a light on the horizon. However, the roll-out of vaccination strategies and their influence on the pandemic are still open problems. In order to compare the effect of various strategies proposed by the World Health Organization and other authorities, a previously developed SEIRS stochastic model of geographical spreading of the virus is extended by adding a compartment for vaccinated people. The parameters of the model were fitted to describe the pandemic evolution in Argentina, Mexico and Spain to analyze the effect of the proposed vaccination strategies. The mobility parameters allow to simulate different social behaviors (e.g. lock-down interventions). Schemes in which vaccines are applied homogeneously in all the country, or limited to the most densely-populated areas, are simulated and compared. The second strategy is found to be more effective. Moreover, under the current global shortage of vaccines, it should be remarked that immunization is enhanced when mobility is reduced. Additionally, repetition of vaccination campaigns should be timed considering the immunity lapse of the vaccinated (and recovered) people. Finally, the model is extended to include the effect of isolation of detected positive cases, shown to be important to reduce infections. 
\end{abstract}

\begin{document}

\flushbottom
\maketitle
\thispagestyle{empty}

\section*{Introduction}
Since the early appearance of SARS-CoV-2 in November 2019 in Wuhan, the world faces a new disease, COVID-19. All governments, without exception, had to quickly devise strategies against this unexpected threat to public health. Meanwhile, the scientific community 
is advancing in the understanding of all aspects of this disease, which has become a pandemic at a fast rate and with a 
strength unknown in contemporary history (i.e. more than 127 million cases and 2.7 million deaths worldwide in 16  months, \cite{Johns:2020}). 
Some countries are going through the effects of a second or third wave of infections while others are still suffering the first. 
As the top priority is to diminish the number of susceptible people, in order to minimize the propagation of the virus, several laboratories started to develop vaccines last year. Many of them have reached promising advances and the first generation of vaccines has been already validated by health authorities in many countries.\cite{Cavaleri:2021, FDA:2020}

At present, various emergency-approved vaccines have reached an efficacy above 90$\%$ in preventing  COVID-19 infections, and more importantly, they protect against serious cases requiring hospitalization and reduce lethality. 

As a consequence, different vaccination plans and strategies have emerged around the world. More than one year after the isolation of the virus \cite{Zhu:2020} some people have already been vaccinated (about 320 million people, i.e. $\sim$ 4.1 $ \%$ of the world population, have received a single dose of vaccine at least, the proportion depending on the country \cite{ourworld:2020}).

Nevertheless, at present vaccine shortage prevails, and international requests are made to the laboratories to increase their vaccine production.
More recently, new variants of SARS-CoV-2 with different mutations were  detected (e.g. in UK \cite{Kirby:2021}, Brazil, SouthAfrica \cite{WHO:2020C}), which are more transmissible than the previously circulating strain  of the virus. 

This  raises questions about the efficacy of the first set of approved vaccines 
against the new strains. 
In any case, this reinforces the need to minimize the propagation of the virus as fast  as possible, since contagion favors the appearance of mutations and eventually also of vaccine-resistant strains. 
 
Due to the novelty of the problem, many open questions remain. A critical point to be considered is the immunity period of time provided by the different developed vaccines. It is not clear yet if the immunity lasts a few months, or if it could reach a year. 
Therefore, even if the problems regarding production, commercialization, distribution logistics, legal aspects,etc.  were solved, the schedule or strategies of vaccination are not closed issues. 

The complexity of the problem deserves to be treated multidisciplinarily. In this sense compartmental models have been useful to describe and predict different disease scenarios \cite{Barrio:2012}.
In particular for countries with large territories and heterogeneous population distribution, such as Mexico and Argentina, geographical spread of the disease must be considered. This family of classical compartmental models can be useful tools only if  a degree of local stochasticity is added \cite{Barrio:2013, Barrio:2020}, accounting  for the social behavior (mobility, familiar relationships, etc.)  \cite{Barreiro:2020}. In this work, we have extended previous approaches by including the possibility of a decrease in the number of susceptible people, as a consequence of a schedule of massive vaccination.

 Concretely, we use a SEIRS (Susceptible, Exposed, Infected, Recovered, Susceptible) stochastic model of geographical spreading of the disease
 \cite{Barrio:2020} and added a compartment to account for vaccinated people, which during the vaccine immunity-period are removed from the susceptible group. 
 We fit the model parameters  describing  the pandemic evolution in Argentina, Mexico and Spain, respectively, and we analyze the effect of various vaccination strategies with different number of stages and timing regarding their application.
 
 These strategies are based on  realistic proposals by the World Health Organization (WHO) and  governments, and account for different social behaviors (as lock-down limitations on mobility e.g.), and also if the  vaccines are applied either homogeneously  or limited to the most densely-populated areas.
 
  Finally, we also included the effect of quarantines on detected positive cases, an effective measure used to reduce infections in many countries, in order to improve the description of the pandemic evolution \cite{Barreiro:2020} and thus assess the  long-term vaccination effects more realistically.
  
 Our analyses focus on three different countries (Argentina, Mexico, and Spain), however, the approach   could be  applied to any country or local jurisdiction that needs to devise a vaccination plan.

In the next section we present the model, then  we exhibit and analyze our results, and finally we discuss them, remarking their relevance in order  to optimize the effect of vaccination plans in the context of vaccine shortages.  

\section*{\label{sec:level2}Model}

The present  model includes   the virus  spreading at two levels: (i) Local dynamics: consisting of a compartmental model whose constant parameters are related to the specific disease agent and the host’s immune response, and (ii) Global dynamics of geographical disease spreading: involving mobility parameters related to social habits of the country and affected, at different times, by different non-pharmaceutical interventions by the  different governments.

\begin{figure}[htpb!]
\centering
\includegraphics[width=0.45\columnwidth]{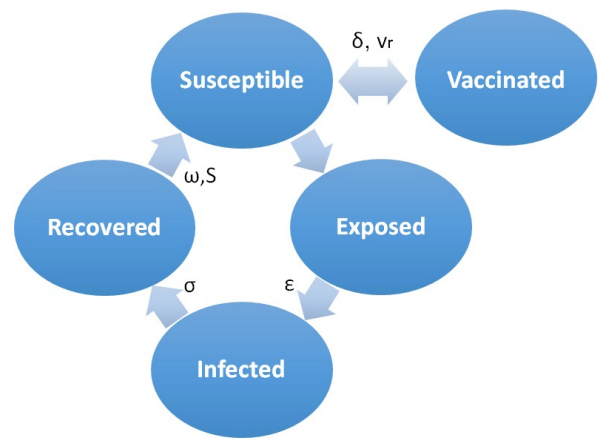}
\caption{Compartment scheme of a SEIRS-V model. $\epsilon$, $\sigma$ and $\omega$ are the latency, infectiousness and immunity periods, $v_r$, $\delta$ and  $S$ are the vaccination rate, the vaccine immunity period and the survival parameter, respectively. }
\label{fig:esquema}
\end{figure}

In order to implement the model, a geographical map of the region of interest is divided into square cells with coordinates $(i,j)$, covering the whole region.  Roads between cities were used to allow short or long distance travelling within the grid.
 Population size ($N$) is assumed to be constant during the simulated period. If life expectancy is given by $L$ and for the mortality we assume an exponential functional form with constant rate $\mu=1/L$, then the  birth rate should be equal to $\mu N$.

Population density heterogeneity is considered by using a matrix whose entries are the actual population densities ($\rho(i,j)$) inside each cell ( at position  $(i,j)$)

The local dynamics within each cell is calculated using a model with four different compartments (see Fig. \ref{fig:esquema}).  These include the susceptible individuals ($S(i,j)$) that might become  exposed ($E(i,j)$) yet not infectious. They   become infected ($I(i,j)$) after the incubation period $\epsilon$ and  remain in that state  for $\sigma$ days. Afterwards   they become recovered ($R(i,j)$) individuals, remain immune for $\omega$ days and close the cycle by  becoming susceptible again according to the survival parameter $S$.
Vaccines are applied to susceptible people according to a vaccination rate $v_r$.  Considering that vaccines arrive in batches, this value is adjusted in each study in order to guarantee the proper administration of all available vaccines in each lot.
It is  assumed that individuals in the vaccinated ($V(i,j)$) compartment have already received as many shots as recommended by pharmaceutics and have developed immunity, thus do not transmit the infection. 

However this immunity lasts $\delta$ days and people become susceptible again after that period.
Vaccines are considered 100$\%$ effective which is a strong assumption. 
All time parameters are constant and dimensionless, expressed in a time scale of one day. Based on these assumptions the model is expressed by the discrete mathematical map detailed in eqs. 1-5 .

\begin{eqnarray} \label{ec:modelo1}
S_{t+1}(i,j) &=& q~[S_{t}(i,j)- G_{t}(i,j)+S q^{\epsilon +\sigma + \omega} G_{t-1-\epsilon -\sigma - \omega}(i,j)+q^{\delta-1} vr_{t-\delta}~ S_{t-\delta}(i,j) - vr_{t}~ S_{t}(i,j)]+\mu N \\
E_{t+1}(i,j) &=& q~[E_{t}(i,j)+ G_{t}(i,j) - q^{\epsilon} G_{t-1-\epsilon }(i,j)]\\
I_{t+1}(i,j) &=& q~[I_{t}(i,j)+ q^{\epsilon} G_{t-1-\epsilon}(i,j)-q^{\epsilon +\sigma } G_{t-1-\epsilon -\sigma}(i,j)]\\ 
R_{t+1}(i,j) &=& q~[R_{t}(i,j)+q^{\epsilon +\sigma } G_{t-1-\epsilon -\sigma}(i,j)- q^{\epsilon +\sigma + \omega} G_{t-1-\epsilon -\sigma - \omega}(i,j)]\\ 
V_{t+1}(i,j) &=& q~[V_{t}(i,j) + vr_{t}~ S_{t}(i,j) - q^{\delta-1} vr_{t-\delta}~ S_{t-\delta}(i,j)]
\end{eqnarray}

 The sum $N(i,j)=S(i,j)+E(i,j)+I(i,j)+R(i,j)+V(i,j)$, is normalized to 1 at t=1. 
 The incidence function $G_t(i,j)=S_t(i,j)\rho(i,j)(1-e^{- \beta I_t(i,j)})$ is based on a Poisson probability distribution, assuming a homogeneous mixing within every cell , and taking into account the population density of the specific cell $(i,j)$. The transmission parameter $\beta$  is constant and does not depend on population density or mobility but it is a characteristic of the specific pathogen. 

It is considered that people sometimes move locally in non-predictable ways. Therefore, randomness is added to  cell dynamics. A local mobility parameter per cell ($0<\nu_L<1$) is compared with a random number from a flat distribution ($r$); if $\nu_L \leq r $  the epidemic proceeds, otherwise  no new infected cases accumulate in that cell due to  low mobility  during that day.

\subsection*{Global dynamics}

Three mobility mechanisms cause geographical spreading of the virus: by displacements to neighbor cells, by long distance travelling, and by seemingly random trips 
causing the spread of the disease to remote areas.

 Virus spreading  to neighbors is often modeled as a diffusion process. An alternative approach is used here by assuming a stochastic process with average cell to cell mobility $0<\nu_n<1$  using a Metropolis Monte-Carlo algorithm. Cells  considered as possible spreaders are defined with $I_t(i,j)>\eta$, $\eta$ being a parameter related to the infectiousness of the disease. For example, considering  $(i,j+1)$ as neighbor cell 
then if $\nu_n$$\leq$$r$  becomes infected then $S_t(i,j+1)=1-\eta$ and $I_t(i,j+1)=\eta$.

Long distance travel takes place by air and land thus  we define a long distance mobility parameter $0<\nu_a<1$.
A Metropolis Monte-Carlo algorithm is run again to define if the  disease is propagated from cell $(i,j)$ to cell $(m,n)$.  The flows of people  between large cities are greater than between small ones. Therefore, in this case $\nu_a\rho(i,j)\rho(m,n)$ is compared to $r$. As before, spreader cells are those where $I_t(i,j)>\eta$.

We assume noise in the transmission process because people make non routinary moves  in a seemingly random way, allowing for the appearance of epidemic outbreaks in unexpected places. Considering this noise as analogous to the “kinetic energy” of the system, if $e^{-1/KT}>r$ and $\rho(i,j)> $T, (with T being a normalized population density threshold), then the disease is started in the cell (i,j) by making $I_t(m,n)=\eta$ and $S_t(m,n)=1-\eta$.

Numerical calculations were performed using maps of population density and road connections for the various countries. (see Supplementary Information: Fig. S1) The size of the grid was of the order of few Km$^2$. Since in this model parameters associated with the pathogen and host’s immune response are independent of population density and of people’s mobility, we used $\beta=0.91$, $\epsilon=1$, $\sigma=14$ and $\omega=140$, as in R. A. Barrio et al. \cite{Barrio:2020}, and N.Barreiro et al. \cite{Barreiro:2020},  for all countries. For simplicity, mobility parameters are 
considered equal $v_a=v_l=v_n=v$. The parameter $v$ changes along time  
according to non-pharmaceutical interventions adopted by each specific government at specific dates.

\section*{Parameter space exploration}

 The present shortage of vaccines worldwide highlights the need to propose coherent and lasting strategies to eradicate the pandemic. The model above allows to simulate and analyze different scenarios and administration schemes, in order to determine the parameters that are most relevant to stop the spread of the disease. 
 
 With this objective in mind we chose three countries as model systems, namely, Argentina, Spain and Mexico. 
These countries were chosen because they allow to compare specific aspects of the model. For instance, Argentina and Spain have a similar populations (45 and 47 million, respectively) but they are spatially distributed in a different way. While Argentina has a large territory with few overpopulated areas surrounded by territories with few people, Spain has a more homogeneous distribution of its population. This leads to quite different connections among cities (flow of travellers).   

Argentina and Mexico are populated quite heterogeneously but they have very different total populations: 
127 and 45 million respectively.

These features are ideal to compare the effect of the model parameters, such as mobility or the increase in the number of vaccines, in different contexts.

In order to analyze their influence 
in the pandemic evolution,  we considered different vaccination strategies by varying a) The total number of vaccines 
administered in different stages according to their availability ; b) the timing between stages; c) the vaccine immunity time, $\delta$,  which is yet uncertain; d) the social behavior (reflected on people´s mobility) once vaccination begins,  and e) the distribution of vaccines, by assuming that the vaccines are applied homogeneously across the country, or vaccinating only in the most densely populated areas.
We assume that the efficiency of the vaccines is 100\%. If the vaccines were less effective, in this model it would be equivalent to considering that the number of administered vaccines was lower. 

At present,
in most of the world the number of vaccines is lower than needed.
Therefore, it is not possible to ensure their availability, according to the plans made by each government, and many delays are expected. Furthermore, there are multiple variables (political, economical and social) that affect social behavior making it difficult to predict future mobility in each country. Because of this, we emphasize that
we do not intend to make specific predictions (although a concrete application of the model would allow it)  but to compare different vaccination scenarios presently under consideration by the corresponding decision makers.

\subsection*{Immunity period and distribution strategies}

Due to constraints on the availability of vaccines in the great majority of countries, different vaccine distribution schemes have been proposed (see a review by
O.J.Wouters et al, \cite{Wouters:2021}). 

To avoid a repetition of the uneven access to vaccines in the H1N1 influenza pandemic of 2009, in April 2020  the World Health Organization announced the creation of the Covid-19 Vaccine Global Access Facility (COVAX), joined already by two-thirds of the world´s countries. To protect the majority of the world´s population against the virus spread and its global impact, COVAX was launched with the initial aim of having 2 billion global vaccine doses available by end of 2021, and a distribution in stages. To ensure equitable access to the world´s largest and most diverse portfolio of safe and effective vaccines, each participant country can request vaccines to cover between 10 and 50$\%$ of its population, though will not receive above 20$\%$ before all other interested countries received that percentage. 

Studies support such a global equitable scheme (see e.g in Chinazzi et al. \cite{Chinazzi:2020}),  predicting a 50$\%$ reduction in the number of global deaths could be obtained if the Covid vaccines were distributed proportionally to each country´s population, instead of reserving two-thirds of the vaccines for higher-income countries.   

Furthermore, considering the expected initial very limited vaccine supply, two initial stages were suggested by the WHO to reach that 20$\%$ coverage of each country´s population: a first stage covering 3$\%$, followed by 17$\%$. Here, we decided to simulate three vaccination stages: those first two WHO-suggested ones, adding a third stage covering 40$\%$ of the population, in order to reach a total of 60$\%$ of the population vaccinated (closer to the expected values for COVID herd immunity).

\begin{figure}[htpb!]
\centering
\includegraphics[width=0.75\columnwidth]{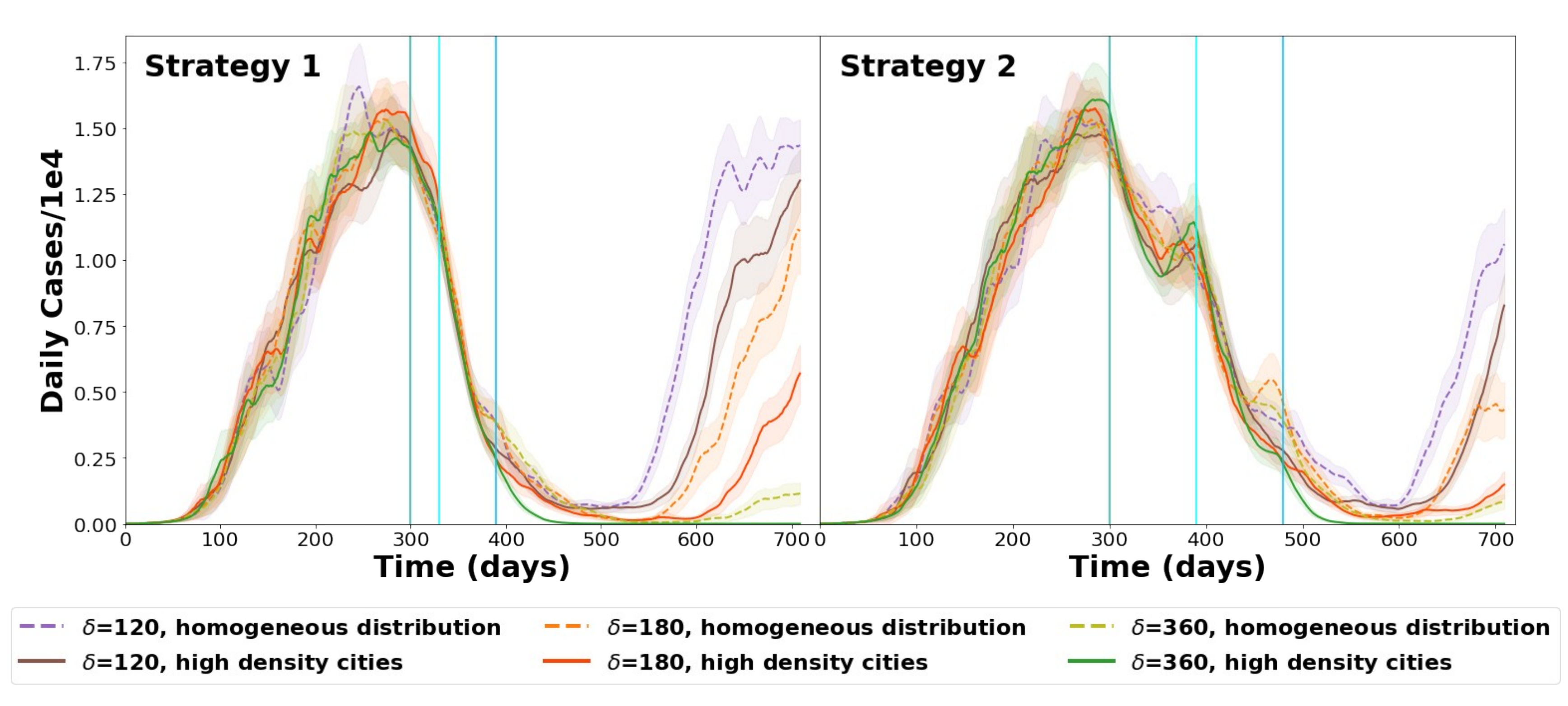}
\caption{Argentina: Dynamics of daily cases, general results for different values of vaccine-induced immunity period ($\delta$). Results for  different periods  between stages and vaccine geographic distributions. Vertical lines indicate the start of each vaccination stage. The shaded areas correspond to the standard error of each curve.}

\label{fig:strategies}
\end{figure}

Since the vaccine immunity period $\delta$ is yet unknown, three values of the parameter were assessed,  varying the time between stages accordingly. Fig.\ref{fig:strategies} shows the results for the case of Argentina. In Strategy 1: the  first stage starts at day 300 after the appearance of the first COVID case, second stage at day 330 and third stage at day 390. In Strategy 2: the first stage starts at day 300, second stage at day 390 and third stage at day 480.  Two different geographic distributions of the vaccines are shown: homogeneous, and  limited to the most densely populated cities. In all scenarios mobility remains constant from day 300 onward.

Figure \ref{fig:strategies} evidences that $\delta$ affects the timing of a new COVID outbreak and also the value of the minimum in the number of daily cases (incidence). 

Incidence  decline depends on the timing between stages, in particular, shorter times between stages are more efficient in lowering the total number of 
cases (see Table S3 in the Supplementary Information). If vaccines are not available for longer periods,  so that the time between vaccination stages is longer, the incidence decline is slower and small rebounds may be observed. Although the new outbreak, up to day 720 seems higher for Strategy 2 (see supplementary information Table S3). 

Our results show that in case of vaccine   shortage, applying doses 
to the most densely populated areas is more effective, since
the total number of cases decreases faster. Furthermore, the minimum number of daily cases reached  is lower than when vaccinating homogeneously in the country. The effect is  clearly observed  for Strategy 1 (further information in Additional Results section of the Supplementary Information).

\subsection*{Vaccination coverage}

\begin{figure}[htpb!]
\centering
\includegraphics[width=0.7\columnwidth]{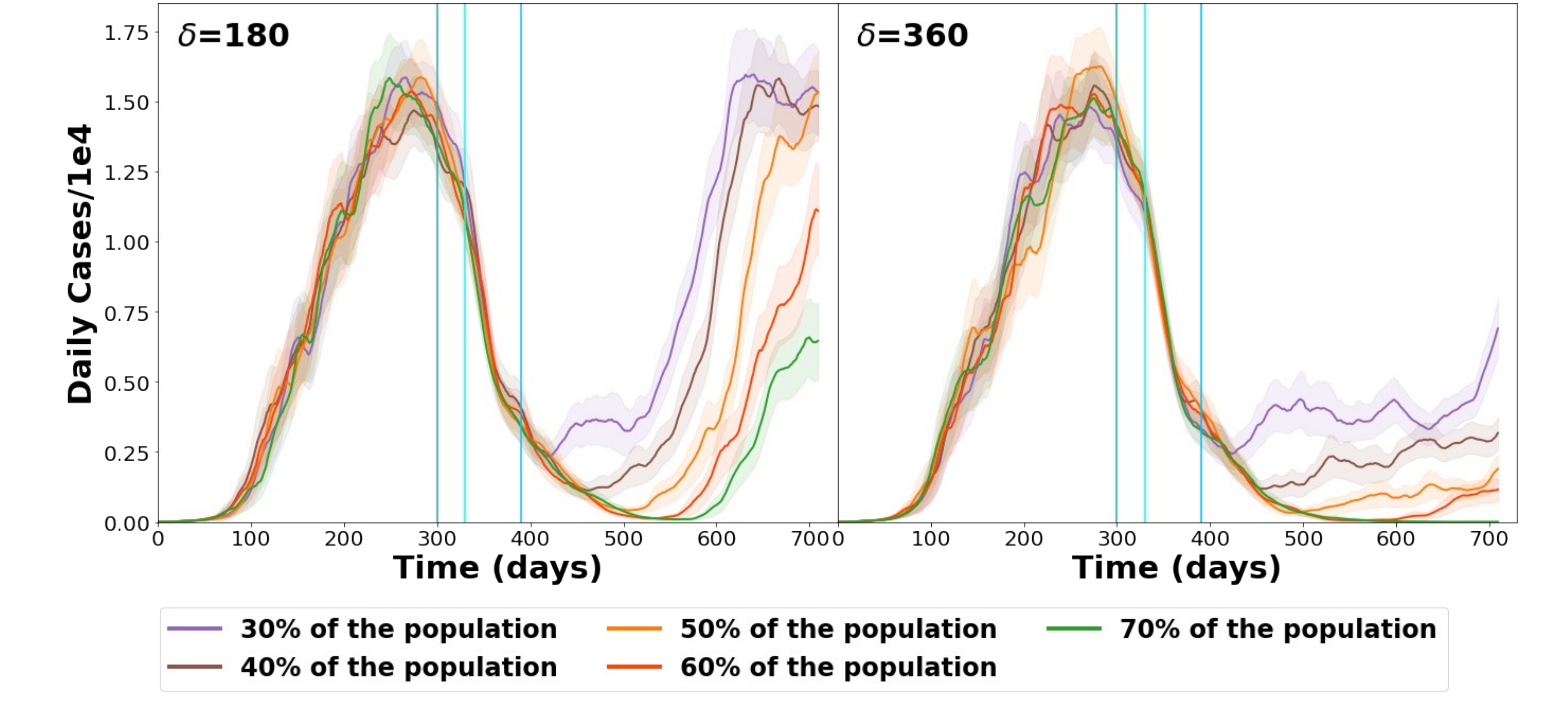}
\caption{Dynamics of daily cases, general results for different vaccination coverage, for two $\delta$ values. Vertical lines indicate the start of each vaccination stage. The shaded areas correspond to the standard error of each curve.}
\label{fig:coverage}
\end{figure}

To further investigate the effect of the total number of vaccines administered, calculations with  Strategy 1 (described above) were performed. An homogeneous distribution of vaccines was simulated while  the number of vaccines applied was changed at the third stage from 10$\%$ to 50$\%$ to obtain a total final coverage of 30$\%$ to 70$\%$. Two values of $\delta$ were used. The results are summarized in Fig. \ref{fig:coverage}.

  For the case of $\delta=180$ days, even with 70$\%$ of the population vaccinated, a second wave is observed. However, if only 30$\%$ of the population is vaccinated, the incidence lowers, but a steady low incidence period is not observed. In all cases, a second wave starts as the number of immune people decreases, even when 70$\%$ of the population is inoculated.
  When 30$\%$, 40$\%$ or 50$\%$ of the population is vaccinated, the number of daily cases on day 720 reaches the same levels observed at the maximum of the first wave.
  
  If $\delta=360$ days, the lowest incidence level reached depends on the percentage of the population vaccinated. However, in most cases a slow but steady incidence increase is observed, leading to the appearance of a second wave.
However, in all cases a slow but continuous increase in incidence is observed, giving rise to the appearance of a new wave.
 Only when vaccinating 70$\%$ of the population the number of daily cases remains very low (approximately 10 cases per day) all thru  to the end of the simulation. 

\subsection*{Impact of social behavior}
\begin{figure}[ht!]
\centering
\includegraphics[width=0.7\columnwidth]{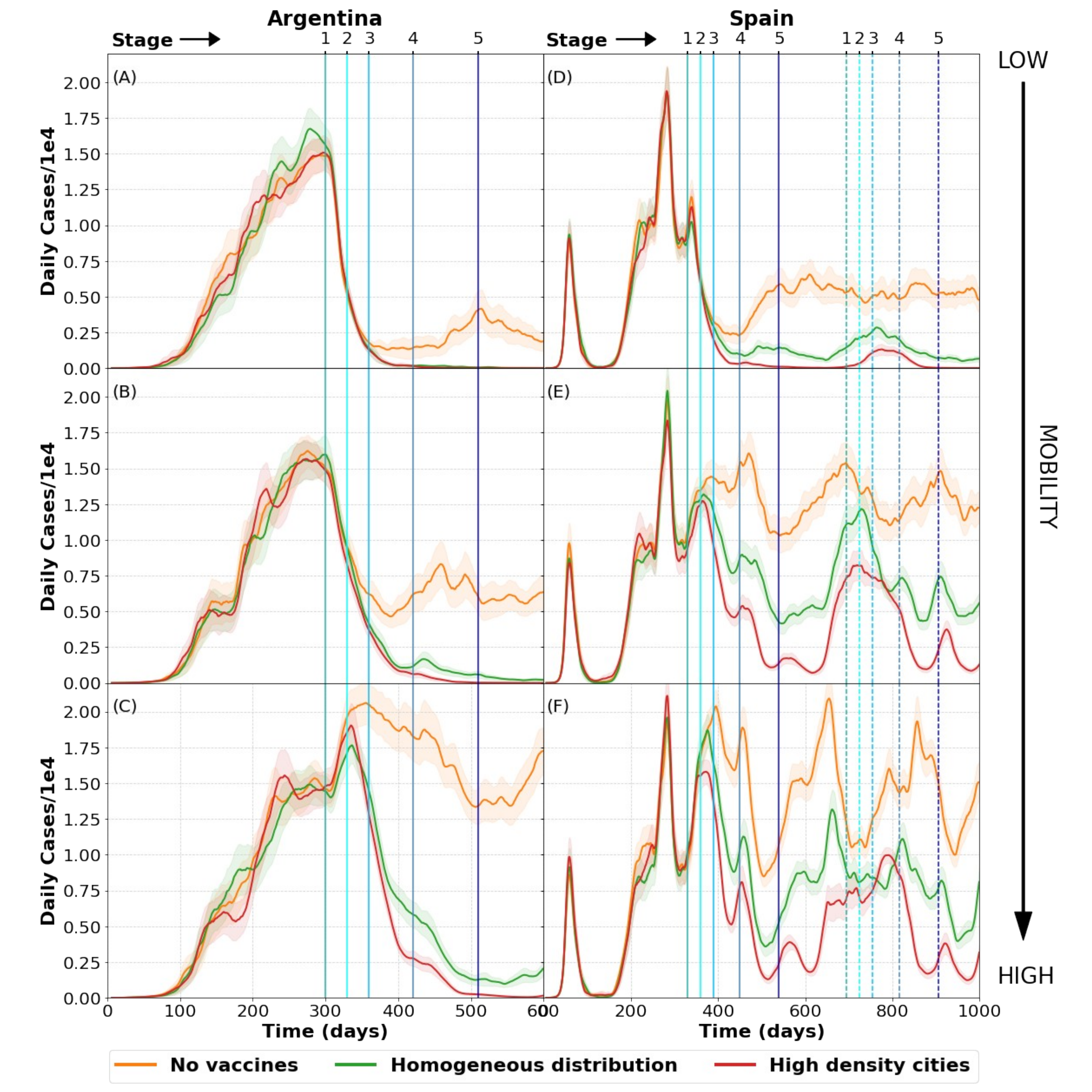}
\caption{Pandemic evolution in Argentina (left) and Spain (right) according to social behavior. Upper figures ((A) and (D))  show the expected number of infected persons in a low mobility scenario; while figures ((B) and (E)) correspond to a medium mobility scenario,  and lower figures ((C) and (F))  to a high mobility scenario. In each graph we consider 3 possible vaccine distributions: None (upper curve), homogeneous,  and vaccines applied only in densely populated areas (lowest curve). 
}
\label{fig:mov}
\end{figure}

Social behavior  is reflected in the mobility parameters used in the model to fit the reported data for a country.  
Fig. \ref{fig:mov} shows the evolution of the pandemic according to different aspects of vaccination and social behavior in Argentina and Spain. All simulations  assume that the vaccines are distributed in 5 stages,  as detailed in the upper panel of Table \ref {tab:table1}. Three curves are compared in each graph:  1)homogeneous vaccine distribution, 2) vaccines applied only in cities with  population density above a threshold (see maps with cities included in Fig. S1 in the Supplementary information), and 3) a baseline without vaccination.

\begin{table}[htpb!]
\centering
\caption{\label{tab:table1} Distribution of vaccines to 62\% of the population in Argentina and Spain  (upper table). Distribution of vaccines to the whole population of Mexico (lower table) }
\resizebox{0.7\linewidth}{!}{
\begin{tabular}{cccc}
 \# of stage & Period   & Argentina vaccines (millions) & Spain vaccines (millons)  \\
\midrule
 1st & January & 0.3 & 0.31   \\
 2nd & February & 5  & 5.2\\
 3rd & March - April & 5 & 5.2 \\
 4th & May - July & 9  & 9.37\\ 
 5th & August onward & 8.7 & 9.06 \\
 \bottomrule
\end{tabular}}
\\
\resizebox{0.4\linewidth}{!}{
\begin{tabular}{ccc}
 \# of stage & Period & Mexico vaccines (millons) \\
\midrule
 1st & February & 0.125   \\
 2nd & March & 17.1  \\
 3rd & April & 13.7  \\
 4th & May & 16.5 	 \\ 
 5th & June onward & rest of the population \\
 \bottomrule
\end{tabular}}
\end{table}
Figure \ref{fig:mov}(A) shows the pandemic evolution in Argentina for the case of low mobility. It simulates  a strict lock-down starting in January 2021. It can be observed , as expected, that the amount of daily cases is  reduced after a few months. In this case there is no difference between vaccine distribution methods as both of them reduce the amount of daily cases to a minimum.
However, due to  economic and social stress, it should not be expected that this low mobility could be sustained  for a long time. Thus other two  situations were modeled, namely with medium and high mobility. 

Figure \ref{fig:mov}(B) shows the evolution of daily cases when social distancing measures are somewhat relaxed. In this case there is a noticeable difference between providing vaccines or not. Moreover,  while a  homogeneous distribution of vaccines seems to drastically diminish  the amount of daily cases, the strategy that reduces the infections to zero is based on  prioritizing highly populated areas . 

Figure \ref{fig:mov}(C) shows the model simulations with high mobility,  representing the case for no social distancing measures at all. In this situation there is a time lag between the application of the vaccines and their effect, observed on the amount of daily cases,  as they keep increasing for a while. 
 
The vaccination of people in highly populated areas is  clearly a better strategy  to reduce the amount of daily cases 
This is because most of the infectious hot spots appear in densely populated areas. From there people  carry the disease to smaller cities and towns. Moreover, when the main  infection sources are turned off, the disease tends to disappear in the whole country. 
 
 From these considerations it can be concluded that good results could be achieved for Argentina by applying a strategy combining social distancing and vaccination, even if only 62\% of the population were immunized, This percentage corresponds to  28 million people, which is the maximum number of vaccines that might be purchased in Argentina during the whole  2021 year, according to the approved national budget (see Fig. S3 in Supplementary Information).

A similar analysis was performed for the case of  Spain modeling  analogous  vaccination strategies  under different representative mobilities (see upper panel in Table \ref{tab:table1}). Although in this case the strategy is  modified in order to vaccinate the same percentage of people at each stage, as it is the case for Argentina (covering 62\% of the total population).


Figures \ref{fig:mov} (D), (E) and (F) show the daily cases in Spain for low, medium and high mobility respectively. There are  three curves in each graph,  representing the different vaccination schemes : no vaccination, homogeneous vaccination, and vaccination  prioritizing the  densely populated cities. All three figures confirm that delivering vaccines in the most densely populated areas is the most efficient method to reduce the number of daily cases. 

However,  the strategy of distributing doses to only 62\% of the population is not as effective in Spain as it is in Argentina for  reducing the daily cases. Mobility has a more important role in the virus spread in  Spain, as the graphs show. This could be understood in terms of geographic extension and intercity connections: While Argentina has a larger territory and many small isolated cities, Spain is more densely populated and very much connected, in average. Therefore,  further outbreaks  spread faster throughout the Spanish territory making it harder to stop the pandemic. This points to the conclusion  that herd immunity is not only related to the amount of vaccines given but also to other characteristics of the country, for instance, geographic extension, connections, social behavior, etc. 

Notice that in Figures \ref{fig:mov} (D), (E) and (F) a second vaccination scheme was included for year 2022 (a span of 720 days after the  first infected case was reported). Since the immunity of the vaccine is assumed to be of 180 days, a rebound of cases is expected in early 2022 due to the loss of vaccine-immunity. Because of this, a periodical immunization scheme should be considered  essential in order to reduce the impact of the pandemic. More examples stressing this point will be shown in the next section.   

\subsection*{Periodical vaccination scheme}

\begin{figure}[ht!]
\includegraphics[width=0.7\columnwidth]{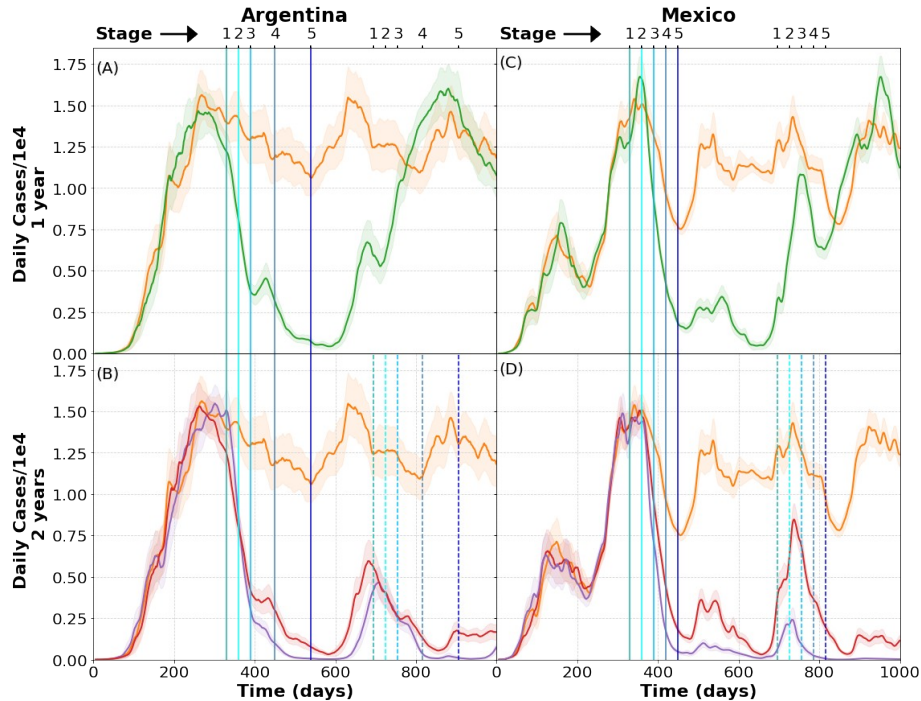}
\centering
\caption{Pandemic evolution in Argentina (A and B) and Mexico (C and D) according to distribution of vaccines. Figures (A) and  (C)  show the daily cases if the vaccines are  applied only during 2021 and the vaccination schemes are not repeated during 2022. In order to compare the results with the cases expected for the regarded mobility, a curve without vaccination (upper curve in each graph) is also included.   Figures (B) and (D) show the effect of adding a new vaccination scheme during year 2022. Here we also confirm that vaccination in the most densely populated areas is an additional strategy to reduce the case count. The shaded areas correspond to the standard error of each curve. }

\label{fig:repeticion}
\end{figure}

 As the pandemic seems to be significantly reduced after an adequate relation between responsible social behavior  and vaccination, a new question emerges. How long will it last? In order to answer this question we propose different scenarios in Argentina and Mexico that are shown in Fig. \ref{fig:repeticion}. The vaccination strategy for Argentina is the same as in table \ref{tab:table1}. For Mexico  we took into account the strategy announced by the government in November 2020. In this case vaccines will be given to the whole population in  progressive steps (like e.g. discussed  by WHO \cite{who:2020b}, or L.Matrajt et al. \cite{Matrajt:2020}).
 
First, Mexico will vaccinate health workers, then, people aged more than 60 years. Next,  people between 50 and 59, after them people between 40 and 49 , and finally, the rest of the population. Taking into account the proportion of the population considered in  each age group, we estimated the vaccination strategy shown in the lower panel of Table  \ref{tab:table1}.  

Figures \ref{fig:repeticion}(A) and (C) show 2 curves: a baseline without vaccines, and homogeneous distribution of vaccines for a scenario where vaccines are given only during 2021. Since in our simulations we assumed that the vaccines provide an immunity period of 180 days, it is expected that the virus will start to spread again and yield a new peak in 2022, reaching the levels of the no-vaccination curve. This behavior is expected to be similar for any vaccine distribution method.

Figures \ref{fig:repeticion}(B) and (D) show 3 curves: a baseline without vaccines and two types of distributions, homogeneous or vaccination only in more densely  populated areas.  In this scenario vaccines are distributed as in Table \ref{tab:table1} during 2021 and 2022 giving place to a significant reduction in the second peak for both countries. It should be noticed that the appearance of this second peak is strongly related to the vaccine immunity period, $\delta$. A longer period would imply a smaller peak, as far as the same strategy is repeated the next year. But if no vaccines are given the second year, the increment in the daily cases will occur anyway at some point in the future. 

As a conclusion and independently of the actual value of $\delta$, it is clear that vaccines should be administered to the population periodically to avoid the long term virus spreading. Again, the application of vaccines in highly populated areas gives  better results than an homogeneous distribution of doses.   

\subsection*{SEIQRS-V model}

Previous studies \cite{Pollan:2020, Stringhini:2020, Salje:2020, Figar:2020} have shown that most countries are capable of detecting only a fraction of the infected cases. The reason being that most people are asymptomatic or present 
mild symptoms \cite{WHO2:2020, Oran:2020} and do not always seek for medical attention \cite{Mizrahi:2020}.

Many articles have emphasized the need to improve infected tracking and contact tracing, as a way to reduce the incidence of the pandemic by isolating those detected as positive  \cite{Keeling:2020, Hellewell:2020, Fraser:2004, Klinkenberg:2020, Gatto:2020}. 
Taking this into account, track and trace policies, along with social distancing and vaccination, should be considered the main pillars to eradicate the pandemic. 

With this in mind, we added a new compartment to our model in order to consider infected people who were isolated. The model with the quarantined (Q) compartment is  described in detail in Barreiro {\it et al.} \cite{Barreiro:2020}. 

In this work we assume that only a fraction $p$ of people who is actually ill is tested and that they get isolated after $ \alpha $ days of being infected. In this way, this $p$ fraction of people stop infecting others reducing the pandemic evolution rate (For more information see Supplementary Information: SEIQRS-V Model section). 

It is important to keep in mind that, in this model, government official data  for infected cases coincide  with the quarantined persons, {\it i.e.} we assume that all the people who tested positive, are in fact isolated and counted as ill by authorities.

In the present work we applied this model to Argentina, Spain and Mexico. As in the SEIRS-V model discussed above, we used the same values for $\beta$, $\epsilon$, $\sigma$ and $\omega$ for all the countries. $p$, $\alpha$ and $S$ where estimated from official data and previous studies. The mobility parameter was fitted in function of time
according to changes in governments policies and social behavior (see Supplementary Information: table S2).

Vaccination strategies used in this case can be found in  Tables \ref{tab:table1} and \ref{tab:table2}. 
In all the cases shown, the same vaccination strategy is repeated in 2021 and 2022. 
The immunity period of the vaccine, $\delta$, was conservatively  fixed at 180 days.  

\begin{table}[htpb!]
\centering
\caption{\label{tab:table2} Distribution of vaccines in Spain (70\% of the population vaccinated until the summer, according to the objectives of the Spanish government \cite{Elmundo:2021}). We delayed the strategy one month, to consider the time needed for actual immunization. The shaded areas correspond to the standard error of each curve.}
\resizebox{0.5\linewidth}{!}{
\begin{tabular}{ccc}
 \# of stage & Period & Amount of vaccines (millons)\\
\midrule
 1st & February & 2.3   \\
 2nd & March & 4.5  \\
 3rd & April & 6.6  \\
 4th & May & 6.6  \\ 
 5th & June & 6.6  \\
 6th & July & 6.6  \\
\bottomrule
\end{tabular}}
\end{table}

\subsection*{SEIQRS-V model comparison}
In this section we compare the results of different vaccination strategies for the three countries considered.

\begin{figure}[htpb!]
\centering
\includegraphics[width=0.85\linewidth]{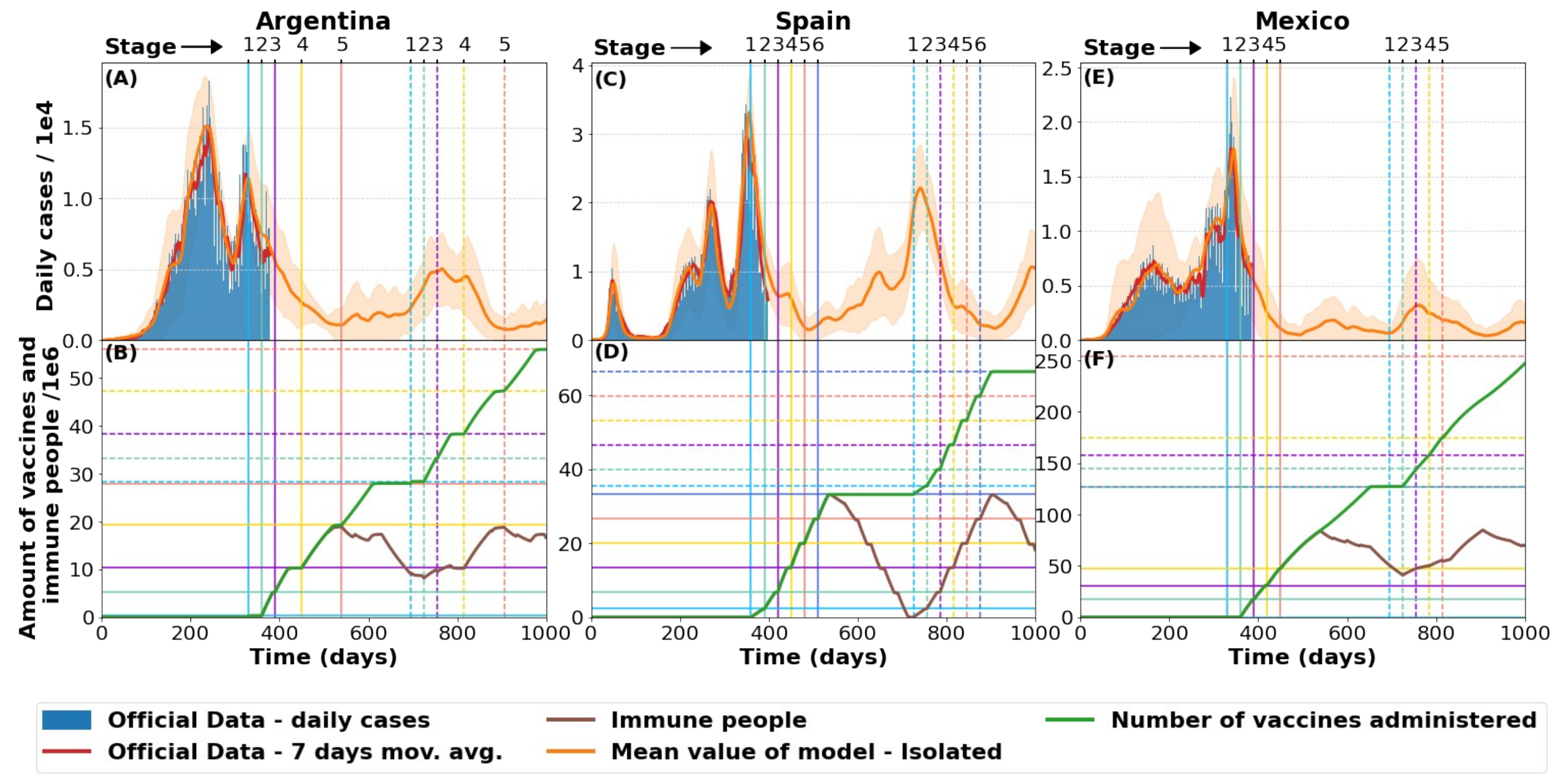}
\caption{SEIQRS-V model for pandemic evolution in Argentina,Spain and Mexico according to distribution of vaccines.Data updated as of March 7, 2021 (A) Model prediction for Argentina, overlapped with 7 day average of actual daily cases. (B) Immunized people and vaccination strategy for Argentina according to distribution of vaccines on Table \ref{tab:table1} . (C) Model prediction for Spain overlapped with 7 day average of actual daily cases. (D) Immunized people and vaccination strategy in Spain according to distribution of vaccines (see Table \ref{tab:table2}).(E) Model prediction for Mexico overlapped with 7 day average of actual daily cases. (F) Immunized people and vaccination strategy in Mexico according to distribution of vaccines (see Table \ref{tab:table1}). Solid and dashed lines respectively  correspond to first and second vaccination year. Horizontal lines in the lower Fig. indicate the available amount of vaccines for each vertical stage (of equal color and aspect).The shaded areas correspond to the standard error of each curve.}
\label{fig:todasseiqrv}
\end{figure}

Figure \ref{fig:todasseiqrv} shows the SEIQRS-V model predictions.
 Day 1 in the figures coincides with the date of the first reported 
COVID-19 case in each country. In the  Johns Hopkins University database \cite{Johns:2020} these dates are March 3rd, February 1st and February 27th for Argentina, Spain and Mexico, respectively.  

For the case of Argentina, mobility was fitted up to day 315 keeping a high value from this moment onward. We also assumed $p=0.1$ and $\alpha=5$. The vaccination plan is the same as in Table \ref{tab:table1} but delayed for 30 days. 

Figure \ref{fig:todasseiqrv} (A) show the number daily cases. One should notice that even with  isolation and vaccination, the model predicts that daily cases will never go below 1000 for high mobility. This implies that it is important to implement a better test and trace mechanism or to vaccinate a higher percentage of the population. 

It should also be remarked that, in this case, we are considering a relatively small value for vaccine immunity (180 days) and this generates the appearance of a second wave around March 2022. As was shown above for the SEIRS-V model, a longer vaccine immunity period delays the appearance of new pandemic waves. 

Figure \ref{fig:todasseiqrv} (B) shows the vaccination strategy as a function of time, and the amount of immune vaccinated people. 
For a $\delta$ value of 180 days the amount of immune people starts descending in November 2021, thus causing a growth of daily cases. When a new batch of vaccines is administered in 2022,  the amount of immune people increases reducing the impact of the new wave during this year. Under the established immunity conditions, shifting the second vaccination periods to November 2021 could prevent the emergence of this new wave.   

Figures \ref{fig:todasseiqrv} (C) and (D) shows the SEIQRS-V model applied to Spain. In this case we used $p=0.2$, $\alpha = 5$ and a high mobility from the day 343 onward. The vaccination strategy is shown in   Table \ref{tab:table2} and applied in 2021 and 2022.  Figure \ref{fig:todasseiqrv} (C) shows the model prediction together with the reported daily cases. If 70\% of the population is vaccinated during the first half of 2021 a strong reduction of daily cases is expected for the summer. Nevertheless, assuming a short immunity period for the vaccine (180 days), the daily cases start growing again generating a new peak around February 2022. This effect could also be understood by looking at Fig. \ref{fig:todasseiqrv}(D). This plot shows vaccinated and immunized people as a function of time.   
Since vaccines are administered within a short time frame, most people lose immunity by the end of the year. This creates a time window with a low number of immune people that allows new outbreaks and the appearance of a new epidemic wave.

Finally, Figures \ref{fig:todasseiqrv} (E) and (F) show the SEIQRS-V model applied to Mexico. In this case we used $p=0.08$, $\alpha = 7$ and a high mobility parameter value from day 325 onward. The vaccination strategy is shown in Table  \ref{tab:table1} and applied in 
2021 and 2022. 

This strategy establishes that vaccines are administered at an approximately constant rate throughout the year. Therefore 
a baseline
of immunized people is created, which causes the reduction of the susceptible group  
 by almost 50 million people. This kind of administration method also reduces the incidence 
 of new pandemic waves making a difference with respect to the cases of Argentina and Spain.

\section*{Final Remarks}

Different vaccination strategies for  the Covid-19 were analyzed using a novel 
epidemiological model with 
stochastic dynamics of 
infections in a geographical region. This approach  has proven to exhibit a reliable long term predictive power   \cite{Barrio:2020, Barreiro:2020}.

The parameters defining the disease, as transmission rate, latency and infectiousness periods, can be adjusted with real data taken during a short time after the first outbreak. Furthermore, the parameters  regulating the spread of the pandemic  in a country have a clear and direct relationship with the measures taken by the authorities and followed by citizens.

This strategy has allowed us  to predict the dynamical behavior of the pandemic in three very different countries, like Argentina, Mexico and Spain. There, we include the suggested vaccination strategies and analyze the results under different scenarios which take into account the  difficulties caused by the current global vaccine shortage.

The mobility parameters of our model allow to simulate different social behaviors (e.g. lock-down interventions, travel restrictions, social distancing), and analyze the effect of different vaccine distribution strategies. One could also analyze the predictions using a strategy based on a homogeneous vaccination of the population, or when prioritizing  the most densely populated areas.

In a context of doses shortage, our  model predicts that the most effective strategy in reducing infections is the vaccination of the most-densely populated areas. 
 
Since the epidemic reappears periodically, we found that vaccination campaigns should be repeated. In order to optimize the effect of these, one should  take into account the immunity period  of the  vaccines to prevent premature future outbreaks. Keeping the number of contagions low is of great importance, as  a large number of infections favors the appearance of mutations and, eventually, of new vaccine-resistant strains. 

Regarding the so called herd immunity, understood as the successful suppression of the epidemic reached when enough people are immune to the virus, previous works estimate its  threshold for SARS-CoV-2 in the   range from 10$\%$ to 70$\%$ or more. The lower values might not hold true because of the assumptions made about people interactions in social networks   \cite{Aschwanden:2020}.
However,  for the case of  COVID-19, we learnt  from our model that it depends on the  number of already infected people and the interplay of  disease and vaccine related-factors.
Among the virus factors we considered the influence of the  immunity time after recovery and the  probability of reinfections. While for  the vaccine-related factors, we explored the effects of the time period of immunity conferred by the vaccine, the percentage of population vaccinated, and the pace of vaccine delivery. 

The probability of infection is affected by logistical, social and demographic factors and it is key to the adopted vaccination strategy.  As was mentioned before, there are huge differences between a scheme based on a demographic homogeneous  vaccine delivery  and one that prioritizes the  most densely-populated areas. 
We found that the vaccination impact is larger in countries with less connected cities, combined with a low mobility.

From modelling concrete vaccination strategies in countries with very different demographic features, 
it can be  clearly concluded that in the case of a limited number of vaccines, herd immunity is not achieved using  vaccination strategies  alone. A combination of mobility 
control, detection and isolation of infected people (and probably social distancing measures) is needed to drastically reduce infections. Such measures have negative socio-economic impacts at the local (and global) level. However, these could be minimized by optimizing the allocation of vaccines.

Some guidelines can be learnt from our model as it was shown and discussed in this work. However, the worldwide vaccination allocation should be approached carefully.\cite{emanuel:2020}  There are  130 nations with 2.5 billion people that have not received a single vaccine yet \cite{mati1}, where the COVID-19  virus with  uncontrolled transmission,  has a perfect playground  to evolve into dangerous variants, even immune to the developed vaccines.

An outbreak there, has the potential to become an outbreak anywhere. In this respect, our model could be extended from one particular country to a continent or even to the whole world  in order to compare  different  vaccine global distribution strategies.

 Despite the real threat, the current  global vaccine allocation is driven by national interest and consequently it is highly asymmetric. High-income countries, representing $16\%$ of the world’s population, purchased more than half of all COVID-19 vaccine doses \cite{mati1} available to date, aside the COVAX agreement. This will extend the time frame of the pandemic and the economical burden worldwide \cite{Currie:2021, Chinazzi:2020, mati2}. Unless  a worldwide vaccine allocation strategy  based  on health and epidemiological needs is implemented, humanity will not be able to  achieve global herd immunity soon enough  and will suffer the  consequences of COVID-19 for much longer.

\bibliography{sample}

\section*{Acknowledgements}
RAB, MN and NLB acknowledge support from The National Autonomous University of Mexico (UNAM) and Alianza UCMX of the University of California (UC), through the project included in the Special Call for Binational Collaborative Projects addressing COVID-19. RAB was financially supported by Conacyt through project 283279. MN was partially supported by CONICET. We appreciate useful discussions with Prof. Yang Quan Chen and collaborators.

\section*{Author contributions statement}
N.L.B., C.I.V., T.G. and R.A.B. conceived the idea. N.L.B. proposed and wrote the model in python, created the maps and did the fitting for each country. N.L.B, C.I.V., T.G. and M.N. conducted the simulations. All authors discussed the results, wrote and reviewed the manuscript.

\section*{Additional information}

\textbf{Competing of interests}\\
The authors declare no competing interests.

\end{document}



\maketitle



\section*{Additional information on methods}

\subsection*{Information to run simulations of the model}

\begin{figure}[ht!]
\centering
\includegraphics[width=0.4\textwidth]{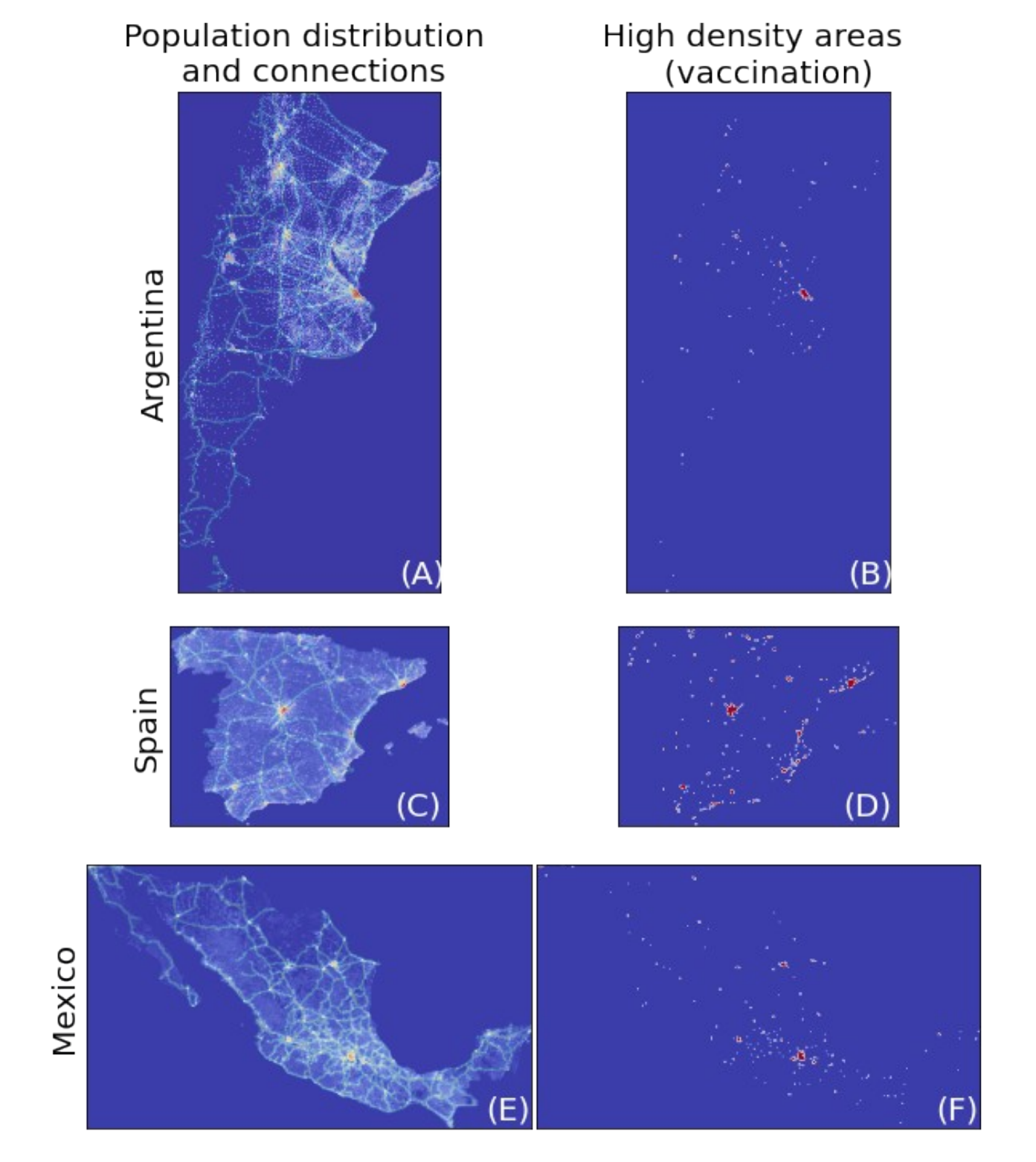}
\caption{Maps used to run the model. Figures (A), (C) and (E) are the population distribution and connection maps used in Argentina, Spain and Mexico respectively to run the code. Figures (B),(D) and (F) exhibit the densely-populated  areas where vaccines are applied.}
\label{fig:mapas}
\end{figure}

As was mentioned before, in order to run the model code, it is necessary to have a population map and the main road connections inside each country. Each map is divided in a grid of square cells in order to consider micro and macrodynamics of the disease. The information from Argentina was obtained from the IGN (Instituto Geográfico Nacional). The population information from Spain and Mexico was obtained from worldpop \cite{worldpop:2018} and the connections were obtained from Google Maps. With the retrieved information we generated the maps from fig \hyperref[fig:mapas]{S1}. In the case of Argentina and Mexico we divided the map in a grid of 7 x 7 $km^2$ cells. The smaller area and higher density of Spain allowed to separate its map in a smaller grid with 5 x5 $km^2$ cells  The three maps on the left, \hyperref[fig:mapas]{S1(A)}, \hyperref[fig:mapas]{S1(C)} and \hyperref[fig:mapas]{S1(E)} show the density maps overlapped with the connections used to run the model. The maps on the right \hyperref[fig:mapas]{S1(B)}, \hyperref[fig:mapas]{S1(D)}, \hyperref[fig:mapas]{S1(F)} represent the areas with higher population used in some vaccination strategies described in the main text.
 
\FloatBarrier

The $\eta$ parameter, which is related with the infectiousness of the disease, was fixed to different values according to population density and grid used in each country. We used $10^{-5}$ , $10^{-4}$ and $5 \, 10^{-6}$ for Argentina, Spain and Mexico respectively. In all the cases it corresponds to starting the disease in a city with around 10 infected people. The survival parameters $S$ used for each country can be found in table \hyperref[tab:table6]{S1}. Final noise and mobility parameters are detailed in the next section.  
\\
\\
\\

\section*{SEIQRS-V Model}

\begin{figure}[ht!]
\centering
\includegraphics[width=0.4\textwidth]{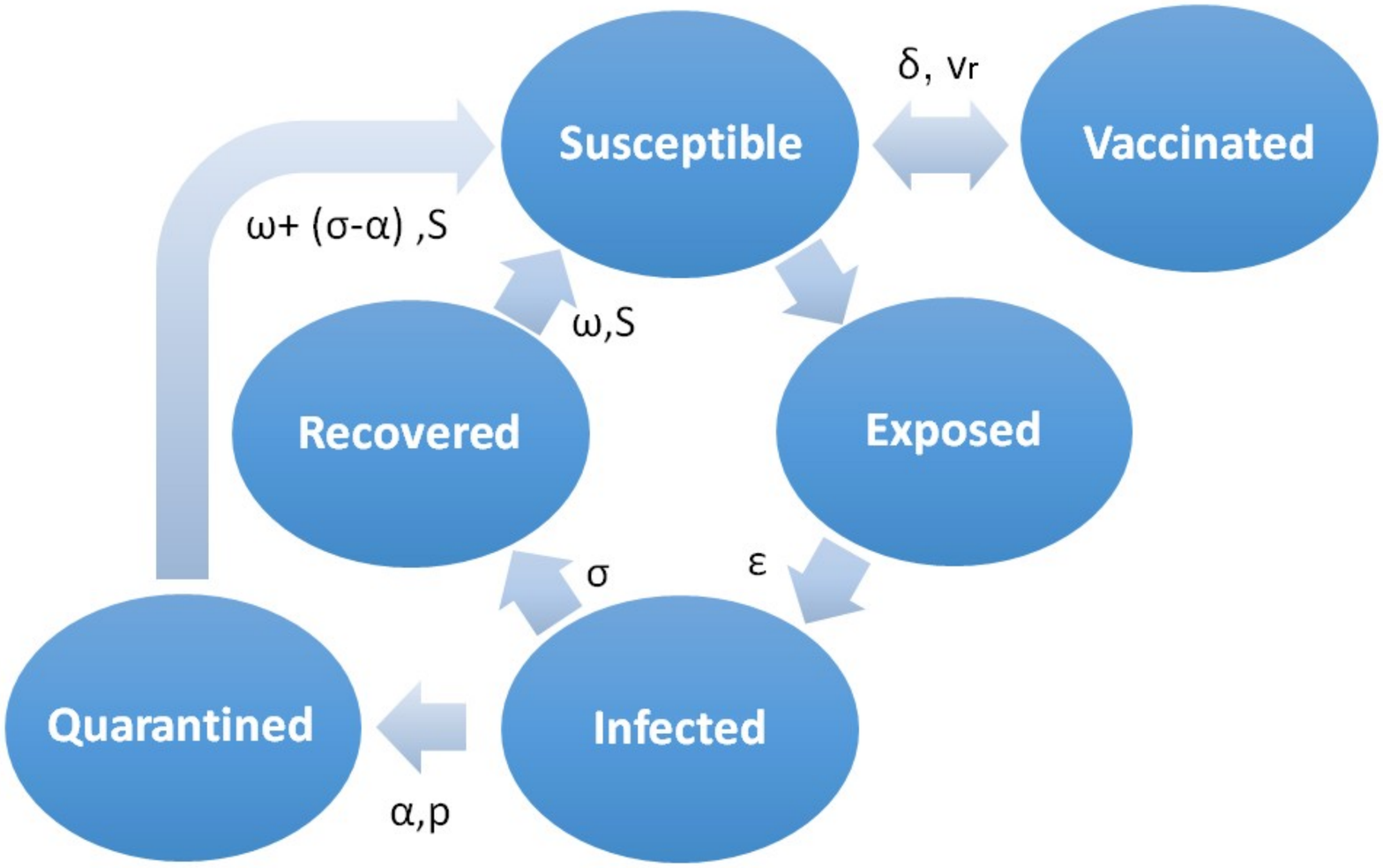}
\caption{Compartment scheme of a SEIQRS-V model. $\epsilon$, $\sigma$ and $\omega$  are the latency, infectiousness and immunity periods, $vr$, $\delta$ and  $S$ are the vaccination rate, the vaccine immunity period and the survival parameter, respectively. $p$ is the fraction of infected isolated people and $\alpha$ the time period from infection to isolation.}
\label{fig:esquema2}
\end{figure}
\FloatBarrier

 In this case the SEIRS-V compartment model was replaced by a SEIQRS-V model where $Q$ stands for quarantined people. Stochastic geographical spread features are the same as those described before. Two new variables were added to consider  the fraction of people discovered and isolated ($p$) and the time period between infection and isolation ($\alpha$). A schematic graph of this model can be seen in figure  \hyperref[fig:esquema2]{S2}. Equations \hyperref[ec:modelo2]{S1-S6} show the addition of this new compartment to the SEIRS-V model. In this case we consider that people tested positive for Covid-19 are not vaccinated until they lose their natural immunity $\omega$.

\begin{align}\label{ec:modelo2}
 S_{t+1} =& (1-\mu)~(S_{t}- G_{t}+ S (1-\mu)^{\epsilon +\sigma + \omega}~ G_{t-1-\epsilon -\sigma - \omega}+(1-\mu)^{\delta-1}~ vr_{t-\delta}~ S_{t-\delta} - vr_{t}~ S_{t} ) +\mu N \tag{S1}\\
 E_{t+1} =& (1-\mu)~(E_{t}+ G_{t} - (1-\mu)^{\epsilon}~ G(t-1-\epsilon )) \tag{S2}\\
 I_{t+1} =& (1-\mu)~(I_{t}+ (1-\mu)^{\epsilon}~ G_{t-1-\epsilon}- (1-p)~(1-\mu)^{\epsilon+\sigma}~ G_{t-1-\epsilon -\sigma }-p~(1-\mu)^{\epsilon+\alpha} ~G_{t-1-\epsilon -\alpha }) \tag{S3}\\
 Q_{t+1} =& (1-\mu)~(Q_{t}+p~(1-\mu)^{\epsilon+\alpha}~G_{t-1-\epsilon -\alpha }-p~(1-\mu)^{\epsilon+\sigma + \omega}~G_{t-1-\epsilon -\sigma - \omega} ) \tag{S4}\\
 R_{t+1} =& (1-\mu)~(R_{t}+(1-p)~(1-\mu)^{\epsilon +\sigma }~G_{t-1-\epsilon -\sigma}- (1-p)~(1-\mu)^{\epsilon +\sigma + \omega}~G_{t-1-\epsilon -\sigma - \omega}) \tag{S5}\\ 
 V_{t+1} =& (1-\mu)~(V_{t} + vr_{t}~S_{t} - (1-\mu)^{\delta-1}~ vr_{t-\delta} ~S_{t-\delta}) \tag{S6}
\end{align}

\FloatBarrier

\subsection*{Model fitting}

The parameter $p$ was estimated from bibliography and local surveys from each country. For instance, a study by Figar et al. \cite{Figar:2020} suggests that only 10\% of the infected are discovered in Argentina. A zero prevalence study in Spain by Pollan et al. \cite{Pollan:2020} indicates that the $p$ parameter should be between 0.1 and 0.3. In our case we were able to fit the data using $p=0.2$. Finally, an ongoing survey from ENSANUT \cite{ensanut:2020} suggest a high presence of COVID-19 antibodies in the Mexican population, implying a small $p$ value. 

\begin{table}[ht!]
\caption{\label{tab:table6} Parameters used in each simulation}
\centering
\begin{tabular}{cccc}
\hline
 Country & $p$ & $\alpha$ & $S$\\
\hline
 Argentina & 0.1 & 5 days & 0.9973 \\
 Spain & 0.2 & 5 days & 0.9973 \\
 Mexico & 0.08 & 7 days & 0.9919 \\
\hline
\end{tabular}
\end{table}
 \FloatBarrier
The value  $\alpha$ was estimated from official data in each country. Assuming that a person can be contagious at least 48 hours before presenting symptoms,  we added 2 days to the time between appearance of symptoms and actual testing. Nevertheless, for small values of $p$, variations in the time between infection and isolation have low impact in the results \cite{Barreiro:2020}. Table \hyperref[tab:table6]{S1} shows the values of $\alpha$ and $p$ used in each case. The table also shows the survival parameter used in each case.
We estimated the dates related with changes in mobility and overall social behaviour from two main sources. The first one was the government intervention dates related with restriction of population mobility and implementation of social distancing policies. This information was acquired from the Oxford stringency index \cite{Thomas:2020}. This index measures the strictness of the government policies taken to face the pandemic on time. The other source was the COVID-19 Community Mobility Reports from Google LLC \cite{google:2020}  Table \hyperref[tab:table7]{S2} shows intervention times and mobility parameters fitted to each country.

 \begin{table}[ht!]
\centering 
\caption{\label{tab:table7} Mobility and noise parameters fitted for each country. Around day 300 all the countries show an increase of mobility because of the holidays}
\centering
\begin{tabular}{llllllllll}
 
  &\textbf{Argentina} & &  & \textbf{Spain} & &  & \textbf{Mexico} & \\
 \hline
  \textbf{Intervention day} &  \textbf{Mobility} & \textbf{Noise} &\vline \textbf{Intervention day} &  \textbf{Mobility} & \textbf{Noise} &\vline  \textbf{Intervention day} &  \textbf{Mobility} & \textbf{Noise} \\
 \hline
 0 & 0.33 & 0.1 &\vline  0 & 0.46 & 0.1 &\vline  0 & 0.73 & 0.1 \\
 23 & 0.135 & 0.1 &\vline  32 & 0.23 & 0.1 &\vline  16 & 0.183 & 0.1 \\
 79 & 0.188 & 0.1 &\vline  39 & 0.06 & 0.1 &\vline  60 & 0.161 & 0.1 \\
 102 & 0.223 & 0.1 &\vline  101 & 0.08 & 0.1 &\vline  92 & 0.319 & 0.3 \\
 166 & 0.356 & 0.1 &\vline  130 & 0.12 & 0.1 &\vline  174 & 0.412 & 0.3 \\
 235 & 0.28 & 0.1 &\vline  143 & 0.2 & 0.1 &\vline  225 & 0.42 &0.5 \\
 285 & 0.51 & 0.1 &\vline  152 & 0.3 & 0.1 &\vline  255 & 0.58 &0.5 \\
 315 & 0.356 & 0.1 &\vline  169 & 0.24 & 0.1 &\vline  305 & 1 & 0.9 \\
 -- & -- & -- &\vline  180 & 0.2 &0.1 &\vline  325 & 0.42 & 0.3 \\
 -- & -- & -- &\vline  209 & 0.18 & 0.1 &\vline  -- & -- & -- \\
 -- & -- & -- &\vline  234 & 0.42 & 0.1 &\vline  -- & -- & -- \\
 -- & -- & -- &\vline  258 & 0.18 & 0.8 &\vline  -- & -- & -- \\
 -- & -- & -- &\vline  301 & 0.37 & 0.3 &\vline  -- & -- & -- \\
 -- & -- & -- &\vline  313 & 0.96 & 0.8 &\vline  -- & -- & --\\
 -- & -- & -- &\vline  343 & 0.54 & 0.8 &\vline  -- & -- & --\\
 \bottomrule 

\end{tabular}
\end{table}        
\FloatBarrier


\subsection*{Code}
The model was developed in python using two libraries: NumPy \cite{numpy} and Pandas \cite{pandas}. Plots were created with Matplotlib \cite{matplotlib}. All the curves were obtained by averaging 100 runs of the model. The model code and the archives used to run simulations will be made available upon request.  

\section*{Additional Results}

\subsection*{ Vaccination shortage and Argentina National Budget for 2021}

 Though the vaccination in Argentina is proceeding at slower pace than planned (around 2.98 million people received at least one dose by 29 March,\cite{ourworld:2020}), due to failure to obtain the vaccines from laboratories at the originally expected rate, taking into account the national budget for the government approved by the Congress in Argentina for year 2021, we can ascertain the effect of the vaccination program.  Concretely, the 2021 national budget has envisaged an expenditure for Covid vaccine purchases  amounting to a minimum of 11 million and a maximum of 28 million vaccines. 

\begin{figure}[ht!]
\centering 
\includegraphics[width=0.4\columnwidth]{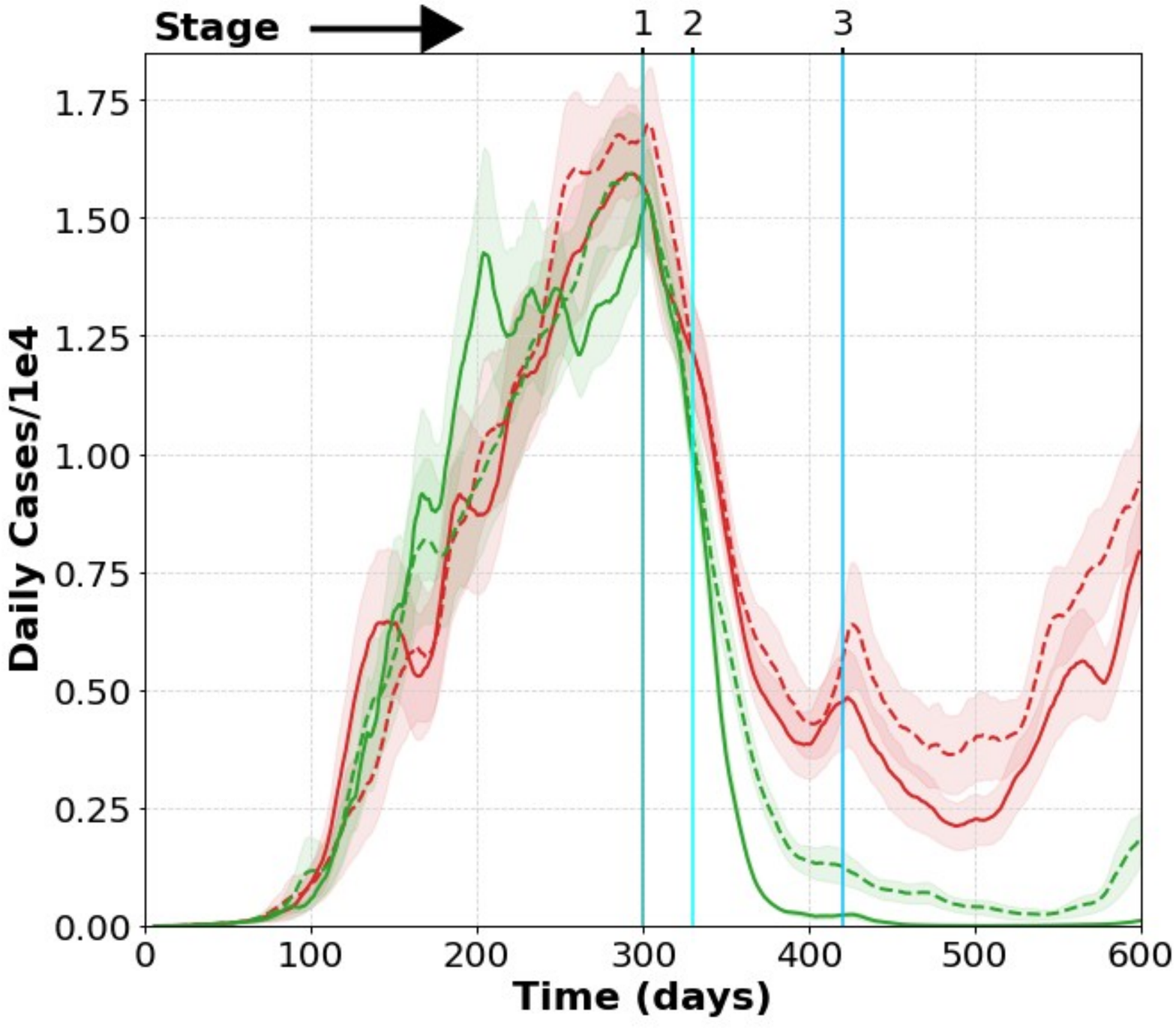}
\caption{Argentina National budget for 2021:   minimal-11 million ( red) and maximal-28 million ( green) total vaccine purchases. Effect of 3-stage vaccination on number of daily infections:  vaccination stages starting at 300, 330, and 420 days (vertical lines) from first infected case (March,3th 2020). Assumed:  immunity time of recovered patients is  140 days, and vaccine immunity time is 180 days.  Solid lines:  cases if vaccination prioritizes densely-populated areas. Dashed lines: cases if vaccines are homogeneously distributed throughout the country. }
\label{fig:arg}
\end{figure}

In Figure \hyperref[fig:arg]{S3}  we show the effect of the minimal and maximal vaccination schemes foreseen by the 2021 national budget, if they were applied in 3 main stages. In the case of the minimal budget: the two first stages give place to an important reduction of the number of infections (to about a third of the main peak of infections), but still an important number of cases would persist during 2021 (a minimum of about 4000 daily cases exists, if vaccines were applied homogeneously in the whole country, or  of 2500 daily cases if vaccination was applied in the most densely-populated cities), with an increase of cases during the last trimester. 

\FloatBarrier

Figure \hyperref[fig:arg]{S3} also shows the effect of the maximal vaccination scheme. In this case, a very important reduction of the number of infections is obtained with the two first vaccination stages: now to about an eighth of the main peak of infections,  reaching a much smaller number of cases afterwards during 2021 (a minimum of about 500 daily cases exists, if vaccines were applied homogeneously in the whole country, and a negligible number of daily cases would be reached if vaccination was applied on the most densely-populated cities), with a small increase of cases during the last trimester. Our results prove that, especially in a context of vaccine shortage, it is clearly advantageous to focus the vaccination on the cities with higher population density in order to more effectively reduce the propagation of the virus. Also, that the maximal vaccination scheme approved in the national budget (reaching about 62 percent of the population) would be required in order to reach a strong reduction of virus propagation in Argentina in 2021.

\subsection*{Additional information on immunity period and distribution strategies}

The results presented in Fig 2 in the main text are clear, however some details are better illustrated in Table \hyperref[tab:table0]{S3}. The upper table shows the accumulated cases at the end of the simulation, equivalent to 720 days of pandemic. Accumulated cases are less when vaccinating high population-density areas in comparison to vaccinating homogeneously throughout the country. For $\delta=120$, the previous result is observed regardless of the timing between stages (Strategy 1 vs Strategy 2); but for $\delta=180$ or $\delta=360$ the total number of cases is lower for Strategy 1 than for Strategy 2. The number of daily cases at the minimum steady incidence period (between days 485 and 540 for Strategy 1, and between days 575 and 630 for Strategy 2), was averaged (lower table). Minimum averages decrease as $\delta$ increases. Notice that for $\delta=360$, using Strategy 1 and vaccinating in high population-density areas, incidences of about 5 cases per day are achieved, making it feasible not only to isolate infected people, but to track their contacts so they can be isolated too.

\FloatBarrier

\begin{table}[ht!] \caption{\label{tab:table0} Total number of accumulated cases (upper table). Minimum number of daily cases (lower table)}
\centering
\begin{tabular}{ccc|cc}
 \multicolumn{1}{c}{}\\
& \multicolumn{2}{c|}{Strategy 1}
& \multicolumn{2}{c}{Strategy 2}\\
\cline{2-5}

 $\delta$ & Homogeneous & High Density  & Homogeneous & High Density  \\
 \hline
 120 & 4,739,832 & 4,084,429  & 4,535,998 & 4,062,800  \\
 180 & 3,758,801 & 3,212,385  & 4,090,868 & 3,695,505  \\
 360 & 3,070,143 & 2,813,648  & 3,779,243 & 3,605,921  \\
\hline
\end{tabular}

\centering
\begin{tabular}{ccc|cc}
 \multicolumn{1}{c}{}\\
& \multicolumn{2}{c|}{Strategy 1}
& \multicolumn{2}{c}{Strategy 2}\\
\cline{2-5}

 $\delta$ & Homogeneous & High Density  & Homogeneous & High Density  \\
 \hline
 120 & 722 & 634  & 979 & 659  \\
 180 & 248 & 203 & 288 &  319  \\
 360 & 270 &  5  & 146 & 137  \\
\hline
\end{tabular}
\end{table}
\FloatBarrier


\bibliography{sample}